\documentclass[letterpaper]{article} 
\usepackage{aaai25}  
\usepackage{times}  
\usepackage{helvet}  
\usepackage{courier}  
\usepackage[hyphens]{url}  
\usepackage{graphicx} 
\usepackage{natbib}  
\usepackage{caption} 
\usepackage{url}            
\usepackage{booktabs}       
\usepackage{amsfonts}       
\usepackage{nicefrac}       
\usepackage{microtype}      
\usepackage{xcolor}         
\usepackage{mathtools} 
\usepackage{booktabs} 
\usepackage{tikz} 
\usepackage{xspace}
\usepackage{graphicx}
\usepackage{subfigure}
\usepackage{cleveref}
\usepackage{enumitem}
\usepackage{amsmath}
\usepackage{algorithmic}
\usepackage[linesnumbered,ruled,vlined]{algorithm2e}

\usepackage{amssymb}
\usepackage{tabularray}
\usepackage{listings}
\usepackage{xcolor}
\usepackage{booktabs}
\usepackage{multirow}
\usepackage{bm}
\usepackage[many]{tcolorbox}
\usepackage{colortbl}

\usepackage{tikz}

\def \VersionWithComments {}
\ifdefined 
\VersionWithComments
\usepackage{marginnote}

\definecolor{codegreen}{rgb}{0,0.6,0}
\definecolor{codegray}{rgb}{0.5,0.5,0.5}
\definecolor{codepurple}{rgb}{0.58,0,0.82}
\definecolor{backcolour}{rgb}{0.95,0.95,0.92}
\definecolor{LightCyan}{rgb}{0.88,1,1}
\definecolor{marine}{rgb}{0.07, 0.04, 0.56}
\definecolor{mygray}{gray}{.9}

\lstdefinestyle{mystyle}{
    backgroundcolor=\color{backcolour},   
    commentstyle=\color{codegreen},
    keywordstyle=\color{magenta},
    numberstyle=\tiny\color{codegray},
    stringstyle=\color{codepurple},
    basicstyle=\ttfamily\footnotesize,
    breakatwhitespace=false,         
    breaklines=true,                 
    captionpos=b,                    
    keepspaces=true,                 
    numbers=left,                    
    numbersep=5pt,                  
    showspaces=false,                
    showstringspaces=false,
    showtabs=false,                  
    tabsize=2
}

\usepackage{newfloat}
\usepackage{listings}
\title{Logic-Q: Improving Deep Reinforcement Learning-based Quantitative Trading via Program Sketch-based Tuning}
\author{
    Zhiming Li\textsuperscript{{\rm 1}}\equalcontrib,
    Junzhe Jiang\textsuperscript{{\rm 2}}\equalcontrib,
    Yushi Cao\textsuperscript{\rm 1},
    Aixin Cui\textsuperscript{\rm 3},
    Bozhi Wu\textsuperscript{\rm 4},
    Bo Li\textsuperscript{\rm 2},\\
    Yang Liu\textsuperscript{\rm 1},
    Danny Dongning Sun\textsuperscript{\rm 5}\thanks{corresponding author}
}
\affiliations{
    \textsuperscript{\rm 1}Nanyang Technological University, Singapore\\
    \textsuperscript{\rm 2}Hong Kong Polytechnic University, Hong Kong, China\\
    \textsuperscript{\rm 3}Chinese University of Hong Kong, Hong Kong, China\\
    \textsuperscript{\rm 4}Singapore Management University, Singapore\\
    \textsuperscript{\rm 5}Peng Cheng Lab, Shenzhen, China
    
    \{zhiming001,yushi002\}@e.ntu.edu.sg,  junzhe.jiang@connect.polyu.hk, 1155209949@link.cuhk.edu.hk, comp-bo.li@polyu.edu.hk, bozhiwu@smu.edu.sg, yangliu@ntu.edu.sg, sundn@pcl.ac.cn

}

\makeatletter
\DeclareRobustCommand\onedot{\futurelet\@let@token\@onedot}
\def\@onedot{\ifx\@let@token.\else.\null\fi\xspace}
\def\eg{e.g\onedot,\xspace} \def\Eg{\emph{E.g}\onedot,\xspace}
\def\ie{i.e\onedot,\xspace} 
 
\def\etc{etc\onedot} 
 
\def\etal{et al\onedot}
\usepackage{bibentry}

\begin{document}

\maketitle

\begin{abstract}
Deep reinforcement learning (DRL) has revolutionized quantitative trading (Q-trading) by achieving decent performance without significant human expert knowledge. Despite its achievements, we observe that the current state-of-the-art DRL models are still ineffective in identifying the market trends, causing them to miss good trading opportunity or suffer from large drawdowns when encountering market crashes. To address this limitation, a natural approach is to incorporate human expert knowledge in identifying market trends. Whereas, such knowledge is abstract and hard to be quantified.
In order to effectively leverage abstract human expert knowledge, in this paper, we propose a universal logic-guided deep reinforcement learning framework for Q-trading, called Logic-Q. In particular, Logic-Q adopts the program synthesis by sketching paradigm and introduces a logic-guided model design that leverages a lightweight, plug-and-play market trend-aware program sketch to determine the market trend and correspondingly adjusts the DRL policy in a post-hoc manner.
Extensive evaluations of two popular quantitative trading tasks demonstrate that Logic-Q can significantly improve the performance of previous state-of-the-art DRL trading strategies.

\end{abstract}
\section{Introduction}
Deep reinforcement learning (DRL) has revolutionized many quantitative trading tasks, \eg~stock trading~\citep{ee2020lstm,nan2022sentiment}, portfolio allocation \citep{guan2021explainable,cui2023portfolio}, order execution~\cite{fang2021universal,zhang2023towards}, \etc. Unlike traditional analytical solutions~\cite{bertsimas1998optimal,almgren2001optimal}, the DRL strategies do not require significant human expert knowledge to design and are able to capture the market’s microstructure automatically~\cite{fang2021universal}. 

\begin{figure}[t] 
\centering 
\includegraphics[width=1.0\linewidth]{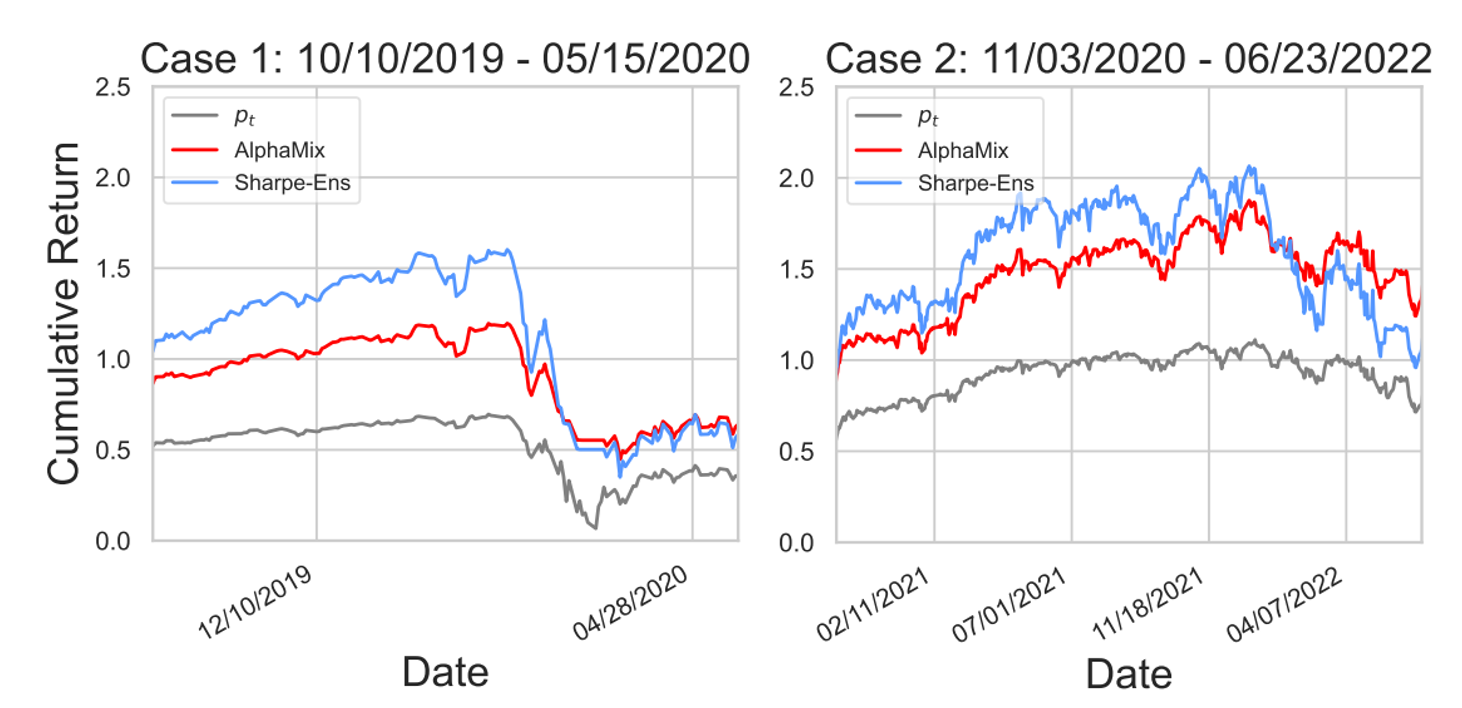} 
	\caption{Illustration of the cumulative return curves of two state-of-the-art DRL stock trading strategies during two market crashes of the US stock market.}
	\label{fig:moti} 
\end{figure}
Despite its great success, previous works present that the DRL strategies are prone to overfitting to spurious noise within the historical context sequence (\ie~training data), which results in their poor performance on testing context sequence~\cite{zhang2023towards} when encountering extreme market conditions.
To achieve more robust and profitable DRL strategies, many recent efforts have been made. \Eg~for the order execution task, Fang \etal~\cite{fang2021universal} propose a novel policy distillation approach based on the teacher-student learning paradigm, which presents better performance on future trading (test) period, and for the stock trading task, Yang~\etal~\cite{yang2020deep,liu2022finrl} introduce a novel ensemble reinforcement learning approach which dynamically selects the best-performing agent policy among three popular DRL policies. This approach has been shown to achieve better risk-adjusted returns during the testing period.

However, even state-of-the-art DRL strategies struggle to accurately identify market trends. This leads to missed trading opportunities and large drawdowns when encountering market crashes.
\Cref{fig:moti} shows the cumulative return curves of two state-of-the-art DRL stock trading strategies (\ie~Sharpe-Ens~\cite{liu2022finrl,yang2020deep} and AlphaMix~\cite{sun2023mastering}) during two market crashes of the US stock market. $p_t$ denotes the Dow Jones Industrial Average (DJIA) market index. Specifically, for both cases, we observe that as $p_t$ descends, both DRL strategies fail to foresee the market crash, and the cumulative return drops significantly. 
A natural idea to improve the performance of the DRL trading strategies is to embed human expert knowledge~\cite{zhou2019devign,brock1992simple} regarding market analysis, as previous works show that the technical indicators of the market are effective in guiding trading decisions~\cite{stoll1990dynamics,lo2000foundations}.
Whereas, the human expert knowledge of market trends is abstract and difficult to quantify because the concrete numeric values associated with market indicators are hard to be manually specified.
\Eg~the following first-order-logic rule represents a straightforward human understanding of a slow decline market trend:
\begin{equation}
\texttt{slow-decline} \leftarrow (\text{VOL}(t) < \alpha) \wedge (\text{DSR}(t) > \beta)
\end{equation}
The downside risk indicator $\text{DSR}(t)$ of the current time step $t$ measures the potential price decrease, and the volatility indicator $\text{VOL}(t)$ measures the degree of variation in the trading price of a market at time step $t$. 
A slow decline in the market trend can be identified when the downside risk exceeds a threshold $\alpha$ while the volatility remains below a threshold $\beta$, indicating a steady downward price movement without significant fluctuations. However, it is hard to specify the detailed value of $\alpha, \beta$ as the market changes over time. To address this and make the best use of the abstract human expert knowledge, in this paper, we adopt the program synthesis by sketching paradigm~\cite{solar2008program} and propose a universal \underline{logic}-guided deep reinforcement learning framework for \underline{Q}-trading, called \textbf{Logic-Q}.
Specifically, Logic-Q adopts a logic-guided model design that proposes using a lightweight, plug-and-play market trend-aware program sketch to embed human expert knowledge of market trends. The program sketch illustrates the overall structure of the abstract human expertise while leaving the concrete numeric details as \emph{hole} to be completed (\ie~parameterized). Once parameterized, the program sketch identifies the market trend for the current time step and adjusts the action probability distribution of the trained DRL policy accordingly, while keeping the model weights frozen. Note that the lightweight design of the program sketch allows it to be efficiently optimized with only a small amount of validation data.
To demonstrate the effectiveness and the generality of Logic-Q, we conduct experiments on two popular quantitative trading tasks (\ie~order execution and stock trading). The results show that the logic-guided design of our framework allows it to significantly improve the performance of the state-of-the-art DRL strategies while being extremely lightweight.

The contributions of our paper are as follows:
 \begin{itemize}
[topsep=2pt,itemsep=2pt,partopsep=0ex,parsep=0ex,leftmargin=*]
  \item We propose a universal logic-guided deep reinforcement learning framework for quantitative trading, called Logic-Q, which can effectively embed abstract human expert knowledge regarding market trends in a logical manner.
  \item Experimental results on two popular quantitative trading tasks demonstrate that Logic-Q can significantly improve the performance of the state-of-the-art DRL strategies while being extremely lightweight.
  \item To the best of our knowledge, it is the first paper to utilize a symbolic model to enhance DRL trading strategies. The general design of Logic-Q may motivate the community to explore its applications in finance further.
\end{itemize}
\section{Preliminary}
\subsection{Order Execution}
Order execution (OE)~\cite{cartea2015algorithmic} refers to the process of carrying out a financial transaction based on a buy or sell order placed by an investor in the financial markets. The goal of the OE is to maximize the revenue by placing the right amount of orders at each timestep. In this work, we follow the formulation of Fang \etal~\cite{fang2021universal}, given the state representation $\mathbf{s}_{t}$ of timestep $t$, a reinforcement learning agent propose a corresponding $a_t \sim \pi(\mathbf{s}_{t})$, $a_t$ is discrete and corresponds to the proportion of the target
order $Q$. And the trading volume to be executed at the next
time is $q_{t+1} = a_t \cdot Q$, and the optimization objective of policy $\pi$ is to maximize the expected cumulative discounted rewards of execution $\underset{\pi}{\arg \max } \mathbb{E}_\pi\left[\sum_{t=0}^{T-1} \gamma^t R_t^{OE}\left(s_t, a_t\right)\right]$ where $\gamma$ is the discount factor. $R_t^{OE}$ is defined as follows:
\begin{equation} 
   R_{t}^{OE}\left(s_t, a_t\right) = \overbrace{a_t\left(\frac{p_{t+1}}{\tilde{p}}-1\right)}^{\text{trading profitability}}-\overbrace{\alpha\left(a_t\right)^2}^{\text{market impact penalty}}
\end{equation}
where $\tilde{p}=\frac{1}{T} \sum_{i=0}^{T-1} p_{i+1}$ is the averaged original market
price of the whole time horizon. And the reward $R_t^{OE}$ specifies that the agent policy should maximize the volume weighted
price advantage (\ie~trading profitability): $a_t\left(\frac{p_{t+1}}{\tilde{p}}-1\right)$ and minimize the market impact (measured with a quadratic penalty): $-\alpha\left(a_t\right)^2$. 

\subsection{Stock Trading}
Stock trading (ST)~\cite{yang2020deep,markowitz1952portfolio} refers to the task that requires a strategy to determine how to adjust a stock portfolio in order to maximize the cumulative return over time. Formally, let $p_{t}\in\mathbb{R}^{D}$ be the share price of $D$ stocks at timestep $t$, $\mathbf{s}_t$ is the state 
representation that contains market information and the trader's remaining balance $b_t$. The RL agent should propose action $\mathbf{a}_t \sim \pi(\mathbf{s}_t), \mathbf{a}_t \in \mathbb{R}^{D}$, which is a vector of actions over $D$ stocks. The action space is defined as $\{-k, \dots, -1, 0, 1, \dots, k\}$, where $-k$ and $k$ represent the maximum number of shares the agent can sell or buy. Finally, the optimization objective of an ST agent policy is to maximize $\underset{\pi}{\arg \max } \mathbb{E}_\pi\left[\sum_{t=0}^{T-1} \gamma^t R_t^{ST}\left(s_t, a_t\right)\right]$, where $R_t^{ST}$ denotes the balance change, formally:
\begin{equation}
    R_{t}^{ST}\left(s_t, a_t\right)=\overbrace{p_{t}^{S}a_{t}^{S}}^{\text{sell change}} - \overbrace{p_{t}^{B}a_{t}^{B}}^{\text{buy change}}
\end{equation}
where $p_{t}^{S}a_{t}^{S}$ represents the balance change due to selling stocks (\ie~sell change), and $p_{t}^{B}a_{t}^{B}$ denotes the balance change due to buying stocks (\ie~buy change).

\begin{figure}[t] 
\centering 
\includegraphics[width=0.49\textwidth]{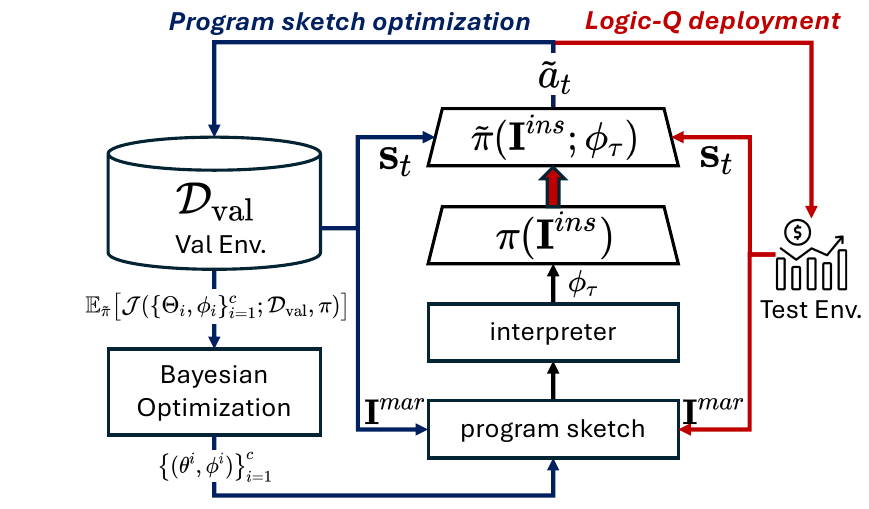} 
\caption{Overview of the Logic-Q framework.} 
\label{fig:overview} 
\end{figure}

\section{Methodology}
In this section, we introduce the details of the proposed program sketch-based tuning (Logic-Q) framework, which effectively combines symbolic market analysis and neural decision-making. The overview of Logic-Q is shown in \Cref{fig:overview}. 
Concisely, we first introduce using a program sketch to embed abstract human expert knowledge for market analysis. It consists of multiple logic conditionals to describe the market trends while leaving numerical details as placeholders to be optimized. We adopt a Bayesian Optimization model to parameterize the program sketch. Once parameterized, it receives the market features $\mathbf{I}^{mar}$ as input and is executed with a program interpreter to determine the market trend $\tau$ of the current time step $t$. Based on the program execution, the program sketch returns the conditional tuning parameters $\phi_{\tau}$. It is used to tune the trained DRL policy $\pi(\mathbf{s}_t)$ accordingly to $\tilde{\pi}(\mathbf{s}_t)$. Due to the symbolic nature of the program sketch, we introduce an effective and efficient optimization approach to optimize its parameters. The following part of this section is organized as follows: we first illustrate the market trend-aware program sketch in detail, then we introduce the program sketch-based tuning approach, and finally, we detail the proposed optimization method of the program sketch.

\subsection{Market Trend-aware Program Sketch}
\begin{figure}[t] 
\centering 
\includegraphics[width=0.44\textwidth]{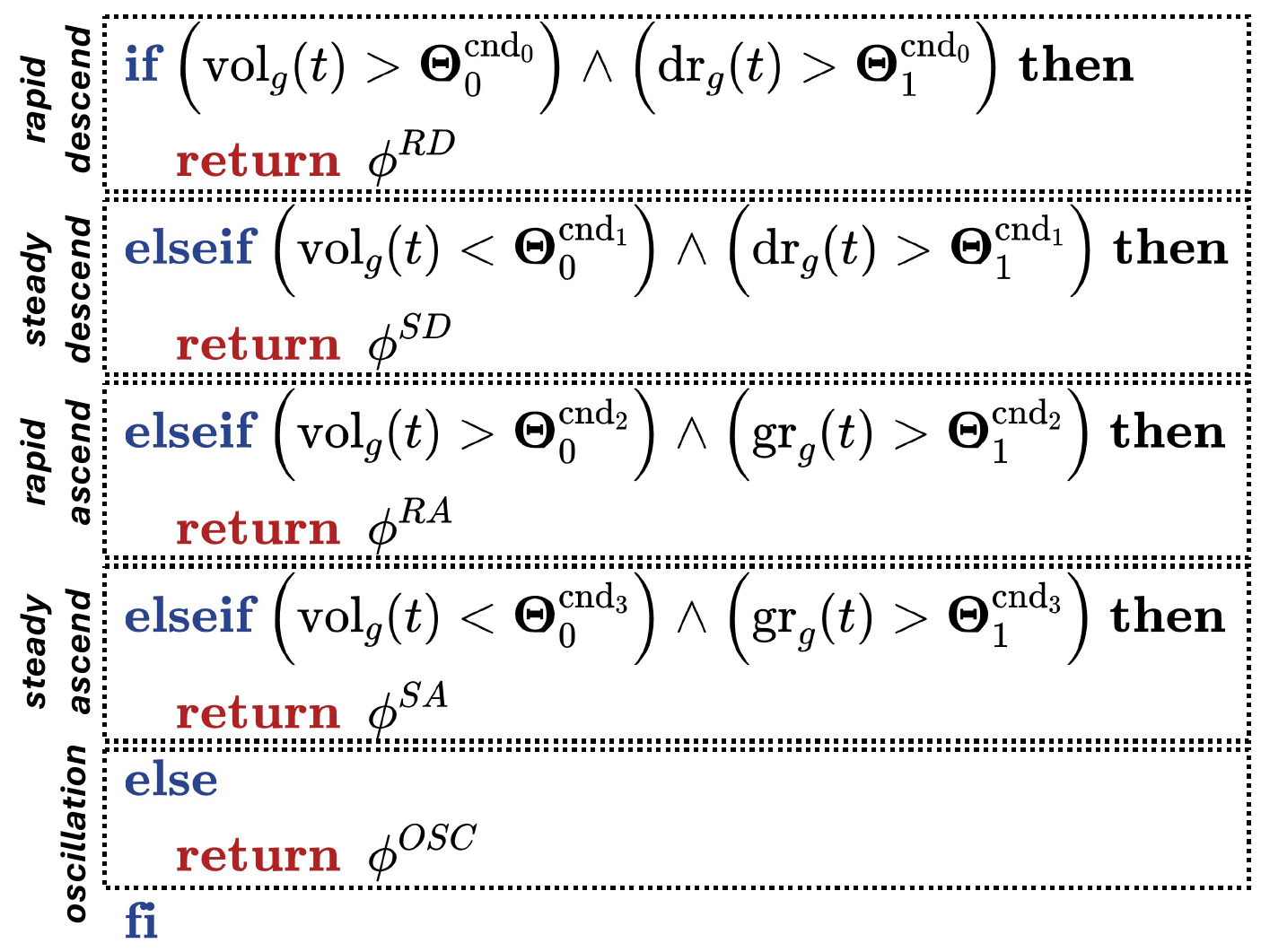} 
\caption{The program sketch used for market analysis.} 
\vspace{-1.5em}
\label{fig:mtps} 
\end{figure}

As aforementioned, the state-of-the-art DRL policies struggle to accurately identify market trends, and the symbolic human expert knowledge regarding market trends is abstract and hard to quantify. In order to effectively combine symbolic market analysis and DRL policies, we adopt the program synthesis by sketching paradigm~\cite{solar2008program,cao2022galois,verma2018programmatically} and propose using a general program sketch to embed abstract human expert knowledge regarding market analysis for DRL policies. 
The detailed program sketch (denoted as $\sigma$) is shown in~\Cref{fig:mtps}. Concretely, it consists of multiple conditional statements, each of them denoting a logical description of a specific type of market trend. The condition of each statement is a boolean expression that logically describes a single market trend. 
In particular, the program sketch describes five market trends: \textit{steady descend}, \textit{steady ascend}, \textit{rapid descend}, \textit{rapid ascend}, and \textit{oscillation}. And it takes market information $\mathbf{I}^{mar}$ as input, which involves three market indicators: $\mathbf{I}^{mar}(t)=\{\operatorname{vol}_g(t),\mathrm{dr}_g(t),\operatorname{gr}_g(t)\}$, where $t$ denotes the current time step. The details of the three technical indicators are as follows:
\begin{itemize}
[topsep=2pt,itemsep=2pt,partopsep=0ex,parsep=0ex,leftmargin=*]
  \item \textbf{volatility} ($\operatorname{vol}_g(t)\in[0,+\infty]$) measures the degree of fluctuation of the market price at the timestep $t$ with a history context window of size $g$, which is calculated as: 
  $\mathrm{vol}_g(t)=\frac{1}{G} \sum_{i=1}^G\left(x_i-\bar{x}\right)^2$, where $x_i$ is the market daily close price and $\bar{x}$ is the mean market price of a time window of size $G$.
  \item \textbf{downside risk} ($\mathrm{dr}_g(t)\in[0,+\infty]$) 
  is defined as\\ $\mathrm{dr}_g(t)=\sqrt{\frac{1}{n} \times \sum_{x_t<\bar{x}}^n\left(\bar{x}-x_t\right)^2}$, where $n$ is the total number of observation below the market mean price $\bar{x}$ and $x_t$ is the corresponding market close price. 
  It measures the variance of the negative return of the whole trading period. We use it as an indicator to reflect the potential loss of the market.
  \item \textbf{growth rate} ($\operatorname{gr}_g(t)\in[0,+\infty]$) measures the percentage of increase over a specified time span. We use it as an indicator to reflect the market's ascending tendency. 
  Formally, \[
\operatorname{gr}_g(t) = 
\begin{cases} 
   \left( \frac{X_{\text{t}} - X_{\text{start}}}{X_{\text{start}}} \right) \times 100\% & \text{if } X_{\text{t}} > X_{\text{start}} \\
   0\% & \text{if } X_{\text{t}} \leq X_{\text{start}}
\end{cases}
\]where $X_{\text{t}}$ is the market close price at timestep $t$ and $X_{\text{start}}$ is the beginning market price in the period of $g$.
\end{itemize}
If a condition is satisfied (\ie~a specific type of market trend $\tau$ is identified), the corresponding consequent will be executed, which returns a tuning parameter $\phi^{\tau}$. 
Take the \emph{steady ascend} market trend for example, if the volatility of the current market is relatively low and the growth rate is high, we can infer that the market is ascending steadily. Whereas, if the volatility and growth rate of the current market are both high, it often denotes a \emph{rapid ascend} market. It is obvious that the definition of lowness and highness is ambiguous. To allow boolean expression with ambiguity, we introduce using the \emph{hole} construct~\cite{solar2008program,cao2022galois}, which represents unspecified scalar/tensor parameters as placeholders to be completed. Therefore, for example, the above-mentioned understanding of the \emph{steady ascend} market trend can be quantified with the boolean expression:$
    \left(\operatorname{vol}_g(t)<\Theta^{\mathrm{cnd}_3}_0\right) \wedge\left(\operatorname{gr}_g(t)>\Theta_1^{\mathrm{cnd}_3}\right)
$, where $\Theta$ and the corresponding consequent $\phi$ are holes to be completed (\ie~parameterized). 
\subsection{Program Sketch-based Policy Tuning}
Given a parameterized program sketch $\sigma_{\{(\Theta^{i}, \phi^{i})\}_{i=1}^{c}}$ (where $c$ is the total number of conditional statements), a program interpreter is used (as shown in \Cref{fig:mtps}) to execute the program, which determines the market trend $\tau$ of the current time step $t$ and returns the corresponding tuning parameter $\phi^{\tau}$. Given the tuning parameter $\phi^{\tau}$ and a trained DRL policy $\pi_{\bm{\theta}}({\bm{s}}_t)$, where ${\bm{\theta}}$ is the model parameter of the DRL agent. We now conduct the post-hoc \emph{program sketch-based policy tuning}, which can be formulated as follows:

\begin{align}  
\tilde{\pi}({\bm{s}}_t,\phi_{\tau}) &= f_{\phi^{\tau}}\circ\pi_{{\bm{\theta}}}({\bm{s}}_t),\\ &\mathbf{where}\ \  \phi^{\tau}
= \sigma_{\{(\Theta^{i}, \phi^{i})\}_{i=1}^{c}}(\mathbf{I}^{mar}({t}))
\end{align}

where $f_{\phi^{\tau}}$ denotes a tuning function that leverages the tuning parameter $\phi^{\tau}$ to adjust a trained DRL policy $\pi_{{\bm{\theta}}}$, ${\bm{s}}_t$ is the state of timestep $t$, $\mathbf{I}^{ins}({t})$ denotes the input features of each instruments, $\tilde{\pi}$ represents the post-tuned policy. 

Specifically, in this work, to demonstrate the generality of Logic-Q, we implement two types of tuning functions for two corresponding reinforcement learning paradigms, namely single-model reinforcement learning~\cite{schulman2017proximal,fang2021universal} and ensemble reinforcement learning~\cite{liu2022finrl,yang2020deep}.
For the single-model RL scenario, where a single agent is used for decision-making: let $\pi_{\bm{\theta}}$ be a trained DRL policy parameterized by ${\bm{\theta}}$, we freeze the DRL model's weight and implement $f_{\phi^{\tau}}\circ\pi_{{\bm{\theta}}}({\bm{s}}_t)$ as logit scaling with Softmax temperature~\cite{agarwala2022temperature}. Formally:
\begin{equation}  
\tilde{\pi}({\bm{s}}_t;\phi_{\tau}) = f_{\phi^{\tau}}\circ\pi_{{\bm{\theta}}}({\bm{s}}_t) = \frac{\mathrm{exp}(\frac{\mathrm{logit}(\pi_{\bm{\theta}}(\mathbf{a}_i \mid {\bm{s}}_t))}{\phi^{\tau}})}{\sum_{j = 1}^{k}\mathrm{exp}(\frac{\mathrm{logit}(\pi_{\bm{\theta}}(\mathbf{a}_j \mid {\bm{s}}_t))}{\phi^{\tau}})}
\end{equation}
Intuitively, $\phi_{\tau}$ is used as the softmax temperature which dynamically adjusts the action probability distribution according to the determined market trend $\tau$ of the current time step $t$. Specifically, for example, if the pretrained DRL policy performs poorly under the rapid descent market trend, the confidence of the DRL policy should be lowered, thus the corresponding $\phi^{\tau}$ would be high: $\phi^{\tau}\in(1,+\infty]$, which smooths the action probability distribution.

For the ensemble reinforcement learning scenario, where multiple models are involved for decision-making: 
let $\Pi=\{\pi_{i}\mid 0<i\leq k\}$ be a set of policies of size $k$, the subpolicies are independently trained. Given the observed state $\mathbf{s}_t$ of timestep $t$, we implement $f_{\phi^{\tau}}\circ\pi_{{\bm{\theta}}}({\bm{s}}_t)$ as the bagging-based ensemble and take the weighted average of the predicted action probability distribution of each policy $\pi_{i}(\cdot \mid s_t;{\bm{\theta}_i}), 1\leq i\leq |\Pi|$, which is as follows:
\begin{equation}
     \tilde{\pi}(s_t;{\bm{\theta}},{\bm{\phi}}^{\tau}) = \sum_{i=1}^{k} {\bm{\phi}}^{\tau}_{i}\cdot \pi_{i}(s_t;{\bm{\theta}_i}),\ 0<i\leq k
\end{equation}
where ${\bm{\phi}}^{\tau}, ||{\bm{\phi}}^{\tau}||_{1}=1$ is used as the weight tensor for combining the 
pretrained subpolicies. Intuitively, the program sketch decides the most suitable combination of the subpolicies under different market trends, which allows us to make the best use of different subpolicies. Besides, note that the program sketch is extremely lightweight. Concretely, for the single-model RL scenario, the program sketch contains only 13 float32 parameters, and for the ensemble RL scenario, the program sketch contains 23 float32 parameters.

\subsection{Program Sketch Optimization}
Finally, we introduce an optimization method for the parameters of the market trend-aware program sketch $\{(\Theta^{i}, \phi^{i})\}_{i=1}^{c}$. Due to the symbolic nature of the program sketch, it cannot be optimized in an end-to-end manner, thus we adopt the Bayesian Optimization~\cite{snoek2012practical} method for the optimization. Concretely, we first specify a task-specific objective function $\mathcal{J}(\phi)$. And since the program sketch only contains an extremely small amount of parameters, we only update the program sketch by maximizing $\mathcal{J}(\phi)$ on a small amount of validation data $\mathcal{D}_{\text{val}}$, formally:
\begin{equation}
    \{\Theta^*_i,\phi^*_i\}_{i=1}^c = \underset{\{\Theta_i,\phi_i\}_{i=1}^c}{\arg \max } \mathbb{E}_{\tilde{\pi}}\left[ \mathcal{J}(\{\Theta_i,\phi_i\}_{i=1}^c;\mathcal{D}_{\text{val}},\pi) \right]
\end{equation}
where $\{\Theta^*_i,\phi^*_i\}_{i=1}^c$ represents the optimal parameters for the program sketch, $\pi$ is the trained DRL policy with weight frozen. Concretely, regarding the two tasks we evaluated in this work, for the order execution task, we use the same optimization objective as the DRL policy (\ie~expected cumulative discounted reward).
For the stock trading task, we optimize the Sharpe ratio~\cite{sharpe1998sharpe} on the validation data to achieve risk-adjusted performance.
Finally, to execute the Logic-Q policy $\tilde{\pi}^{*}$ on the test (trading) data $\mathcal{D}_\text{test}$ (as shown in the right part of \Cref{fig:overview}), we conduct decision-making by executing the optimized program sketch to determine the market trend of the current time step $t$, then we conduct program sketch-based policy tuning on the trained DRL strategy to adjust the neural policy accordingly and we sample the action from the action probability distribution of the post-tuned policy:
$\tilde{\mathbf{a}}_{t} \sim \tilde{\pi}^{*}(\mathbf{s}_{t}), t \in \mathcal{D}_\text{test}
$
.
\section{Experiments}
\begin{figure*}[!t]
\centering     
\subfigure{\label{fig:oe1}\includegraphics[width=0.32\textwidth]{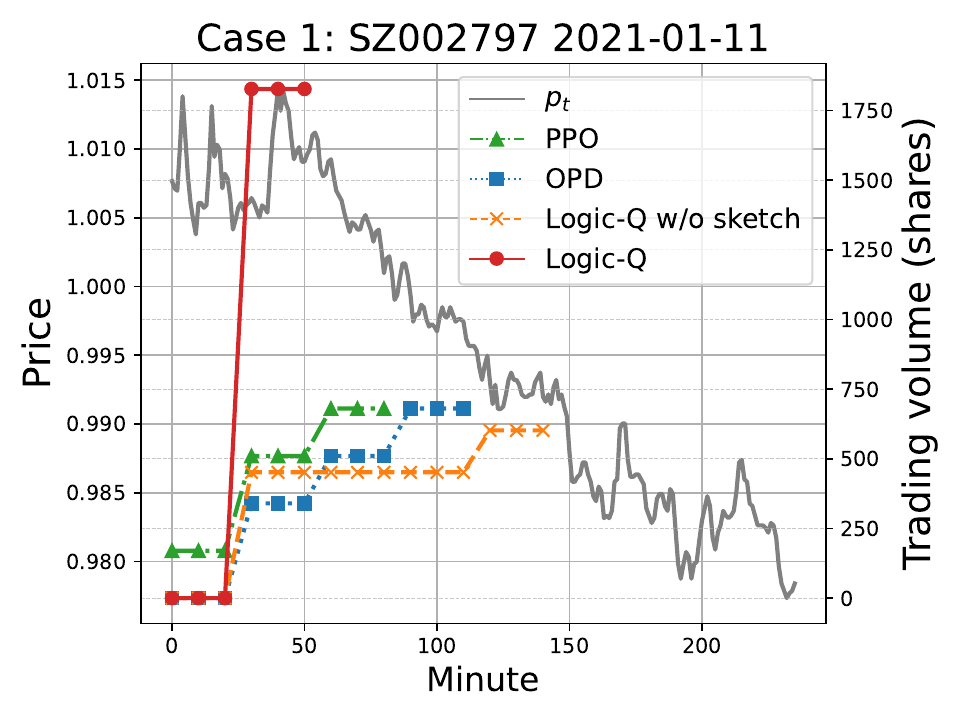}}
\subfigure{\label{fig:oe2}\includegraphics[width=0.32\textwidth]{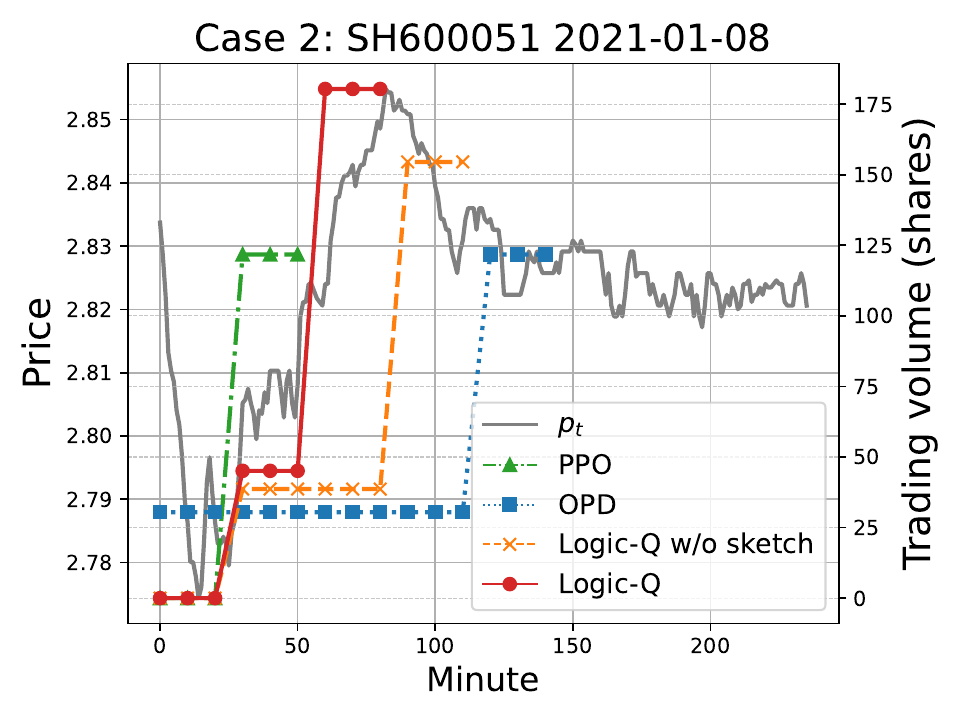}}
\subfigure{\label{fig:oe3}\includegraphics[width=0.32\textwidth]{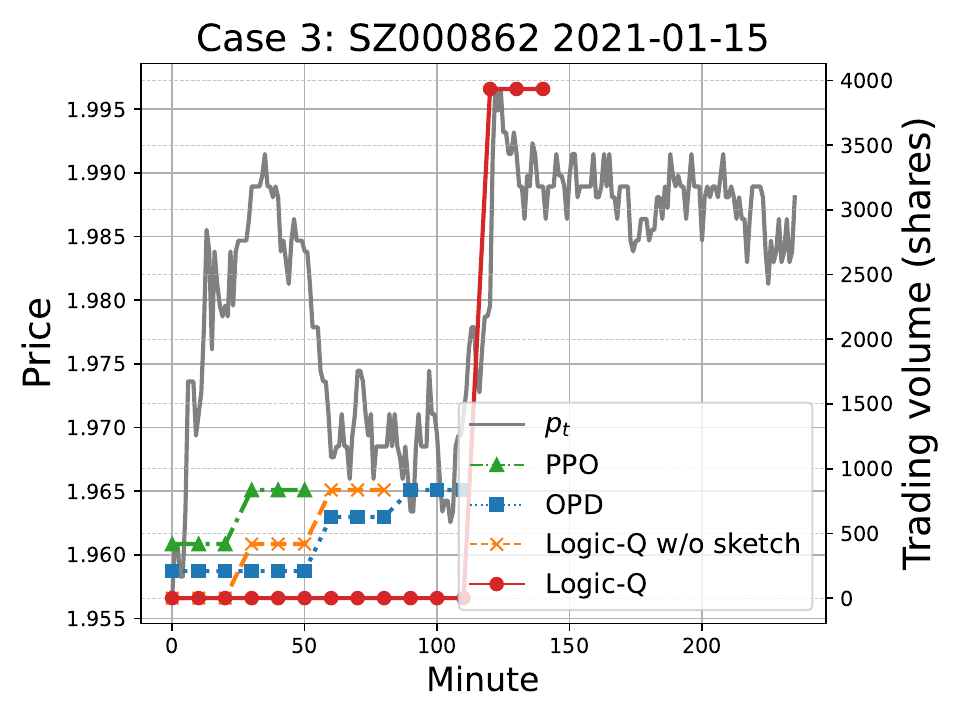}}
 \caption{Case studies of the execution details of different methods.
 }
\label{fig:oe_cs}
\vspace{-1em}
\end{figure*}
To evaluate the effectiveness of the Logic-Q framework, we conduct experiments on two popular quantitative trading tasks. Concretely, we evaluate the Logic-Q's performance under the single-model reinforcement learning setting on the order execution task; and we evaluate the performance under the ensemble reinforcement learning setting on the stock trading task.  In the remainder of this section, we first introduce the detailed experimental setup, and then we 
answer three research questions (RQs) to lead our discussion, which are as follows: \textbf{RQ1:} Can Logic-Q effectively improve a single-model reinforcement learning policy's performance? \textbf{RQ2:} Can Logic-Q effectively improve an ensemble reinforcement learning policy's performance? \textbf{RQ3:} How well does the program sketch's market trend identification align with human expertise?


\subsection{Order Execution Setup}
\subsubsection{Datasets \& Training}
We conduct all experiments on the historical transaction data of the stocks of the China A-shares market provided by Fang et al.~\cite{fang2021universal}. The dataset consists of minute-level intraday price-volume market data of CHINA SECURITIES INDEX 800 (CSI 800) constituent stocks and the order amount of each instrument of each trading day. The results of the evaluated methods are all averaged over 5 random seeds (same for the stock trading task).\footnote{Please refer to the supplementary materials for the detailed experimental setup.}
\begin{table}[!t]
\centering
\resizebox{0.48\textwidth}{!}{
\begin{tabular}{ccccc} 
\toprule
\multicolumn{1}{c}{Method}   & GLR                                   & POS         & PA~

 (‱)                             & ARR

 (\%)     \\ 
\hline
TWAP & 0                                     & 0           & 0                                     & 0              \\
VWAP    & 0.93                                  & 0.53        & 3.40                                  & 8.94           \\
PPO        & 0.91 ± 0.01                           & 0.53 ± 0.01 & 3.27 ± 0.25                           & 8.59 ± 0.63    \\
OPD       & 0.91 ± 0.01                           & 0.52 ± 0.01 & 3.41 ± 0.22                           & 8.97 ± 0.56    \\
OPD (Aug)               & 0.90 ± 0.00                          & 0.51 ± 0.00 & 3.07 ± 0.21                           &   8.15 ± 0.55  \\	
Logic-Q w/o sketch               & 0.92 ± 0.01                           & 0.52 ± 0.01 & 3.50 ± 0.29                           & 9.22 ± 0.73    \\	
\rowcolor{mygray}
\textbf{Logic-Q (Ours)}          & \textcolor{red}{\textbf{0.96}} ± 0.00 & \textcolor{red}{\textbf{0.55}} ± 0.00
 & \textcolor{red}{\textbf{4.33}} ± 0.25  & \textcolor{red}{\textbf{11.53}} ± 0.63      \\
 \hline
Improvement over SOTA        & \textcolor{marine}{{\textbf{3.2\%}}}                          & \textcolor{marine}{{\textbf{3.8\%}}}  & \textcolor{marine}{{\textbf{23.7\%}}}                          & \textcolor{marine}{{\textbf{25.1\%}}}    \\
\bottomrule
\end{tabular}
}
\caption{The comparative results of different methods on the order execution task.}
\label{tab:oe}
\end{table} 

\subsubsection{Compared Methods}
We compare our method with two traditional financial
model-based methods (\ie~Time-Weighted Average Price (TWAP)~\cite{bertsimas1998optimal}, Volume-Weighted Average Price (VWAP)~\cite{kakade2004competitive}), and two state-of-the-art DRL-based order execution methods, namely PPO~\cite{lin2021end} and OPD~\cite{fang2021universal} (refer to the supplementary materials for the details of these methods).


\subsubsection{Evaluation Measurements} 
We evaluate the OE strategies' performance with 4 different measurements:
(1) Price Advantage (PA): PA measures the relative gained revenue of a trading strategy compared to a baseline price, (2) Additional Annualized Rate of Return (ARR): ARR measures the additional annualized rate of return brought by an order execution strategy compared to the TWAP strategy, (3) Gain-loss Ratio (GLR): GLR is a metric that compares the average gain from winning trades to the average loss from losing trades over a trading period, (4) Positive Rate (POS): POS measures the positive rate of PA across all orders over a trading period. 

\subsection{Stock Trading Setup}
\subsubsection{Datasets \& Training} 
For our stock trading analysis, we conducted experiments across three diverse financial markets: the United States stock market, the Hong Kong stock market, and the cryptocurrency market. We use the public end-of-day trading dataset collected by Yahoo Finance\footnote{\url{https://github.com/yahoo-finance}}. For our trading stock pools, we selected the Dow Jones 30 constituent stocks for the US market following Yang \etal~\cite{yang2020deep} and Hang Seng China 50 Index constituent stocks as the trading stock pool. In the cryptocurrency market,we chose ten high-volume cryptocurrencies paired with USD: BTC, ADA, FIL, ETH, LTC, BNB, EOS, ETC, LINK, and BCH. We evaluate Logic-Q and the baselines under both the cash trading and the margin trading scenario. 
For the margin trading setting, we borrow funds equal to the total value of our account, creating a 1:1 loan-to-value ratio. The borrowing amount is adjusted every three months to maintain the same leverage ratio and to repay the interest incurred from borrowing.
\begin{table*}[!t]
\centering
\resizebox{0.85\textwidth}{!}{
\begin{tabular}{cccccc|ccccc} 
\toprule
\multicolumn{11}{c}{US stock market}                                                                                                                                                                                                                                                                                                                                                                                                                       \\ 
\toprule
\multirow{2}{*}{Methods}                 & \multicolumn{5}{c|}{Cash Trading}                                                                                                                                                           & \multicolumn{5}{c}{Margin Trading}                                                                                                                                                                                \\ 
\cmidrule{2-11}
                                         & AR                       & CR                   & AV                   & MD                    & SR                              & AR                       & CR                    & AV                   & MD& SR                            \\ 
\midrule
$p_t$                                    & 9.3\%                              & 91.7\%                               & 18.9\%                             & -38.0\%                             & 0.564                               & 9.3\%                              & 91.7\%                                & \textcolor{red}{\textbf{18.9\%}}                             & -38.0\%                                                   & 0.564                              \\
DDPG                                     & 9.3\% & 82.7\% & 19.6\% & -31.7\% & 0.585 & 15.0\% & 164.8\% & 36.2\% & -49.4\% & 0.587 \\
A2C                                      & 9.7\% & 94.7\% & 17.2\% & -25.2\% & 0.599 & 21.4\% & 303.3.\% & 26.7\% & -31.9\% & 0.851 \\
PPO                                      & 11.5\% & 121.7\% & 16.9\% & -22.7\% & 0.807 & 10.3\% & 105.6\% & 26.8\% & -27.9\%& 0.549 \\
AlphaMix                                      & 14.2\% & 169.7\% & 20.3\% & -33.1\% & 0.816 & 23.4\% & 372.6\% & 29.9\% & -34.3\% & 0.863 \\

Sharpe-Ens                                 & 10.3\% & 124.2\%& 33.7\% & -43.8\% & 0.509 & 22.5\% & 332.9\% & 40.2\% & -49.5\%& 0.736 \\
AlphaMix (Aug)                                      & 15.4\% & 181.9\%& 20.2\% & -38.1\% & 0.831 & 21.9\% & 370.5\% & 23.2\% & -37.0\%& 0.786 \\
Sharpe-Ens (Aug)                                 & 11.4\% & 131.2\% & 30.4\% & -52.7\% & 0.548 & 21.3\% & 318.8\% & 36.6\% & -50.8\% & 0.662 \\
Logic-Q w/o sketch                           & 11.1\% & 125.0\% & 16.7\% & -27.9\% & 0.756 & 24.8\% & 384.2\% & 53.6.2\% & -68.4\% & 0.716 \\

\rowcolor{mygray}
\textbf{Logic-Q (Ours)}                        & \textcolor{red}{\textbf{18.1\%}} & \textcolor{red}{\textbf{239.6\%}} & \textcolor{red}{\textbf{16.5\%}} & \textcolor{red}{\textbf{ -22.3\%}} & \textcolor{red}{\textbf{1.097}} & \textcolor{red}{\textbf{32.9\%}} & \textcolor{red}{\textbf{697.9\%}} & 27.1\% & \textcolor{red}{\textbf{ -26.1\%}} & \textcolor{red}{\textbf{1.175}} \\ 
\hline
Improvement oer SOTA                       & {\textcolor{marine}{\textbf{17.5\%}}}                    & {\textcolor{marine}{\textbf{31.7\%}}}                & {\textcolor{marine}{\textbf{1.2\%}}}                 & {\textcolor{marine}{\textbf{1.8\%}}}       & {\textcolor{marine}{\textbf{32.0\%}}}               & {\textcolor{marine}{\textbf{32.6\%}}}                 & {\textcolor{marine}{\textbf{81.7\%}}}                                       & {\textbf{-}}                 & {\textcolor{marine}{\textbf{6.5\%}}}      & {\textcolor{marine}{\textbf{36.2\%}}}                                                  \\

\hline
\multicolumn{11}{c}{HK stock market}                                                                                                                                                                                                                                                                                                                                                                                                                       \\ 
\hline
\multirow{2}{*}{Methods}                 & \multicolumn{5}{c|}{Cash Trading}                                                                                                                                                           & \multicolumn{5}{c}{Margin Trading}                                                                                                                                                                                \\ 
\cmidrule{2-11}
                                         & AR                       & CR                   & AV                   & MD                    & SR                              & AR                       & CR                    & AV                   & MD                                          & SR                             \\ 
\midrule
$p_t$                                    & -0.1\%                              & -5.4\%                               & 21.3\%                             & -55.7\%                             & 0.069                               & -0.1\%                              & -5.4\%                                & \textcolor{red}{\textbf{21.3\%}}                             & -55.7\%                                                   & 0.069                              \\
DDPG                                     & 0.5\% & 3.7\% & 20.2\% & -52.9\% & 0.131 & 0.8\% & 20.7\% & 57.9\% & -81.6\% & 0.346 \\
A2C                                      & 4.9\% & 35.9\% & 18.1\% & -42.6\% & 0.347 & 4.9\% & 41.2\% & 33.8\% & -61.3\%& 0.352 \\
PPO                                      & 4.2\% & 39.4\% & 24.1\% & -51.4\% & 0.332 & 1.8\% & 15.1\% & 39.0\% & -68.6\% & 0.245 \\
AlphaMix                                      & 5.3\% & 46.4\% & \textcolor{red}{\textbf{19.4\%}} & -40.7\% & 0.473 & 16.7\% & 223.0\% & 34.1\% & -51.9\% & 0.727 \\

Sharpe-Ens                                 & 0.8\% & 8.0\% & 38.3\% & -75.4\% & 0.184 & 11.6\% & 125.5\% & 42.7\% & -56.8\% & 0.483 \\
AlphaMix (Aug)                                      & 5.3\% & 49.9\% & 20.3\% & -38.7\% & 0.465 & 19.2\% & 234.6\% & 34.5\% & -47.8\%& 0.926 \\
Sharpe-Ens (Aug)                                & 1.2\% & 9.0\% & 29.7\%& -68.9\% & 0.287& 14.6\% & 131.7\% & 41.6\% & -63.7\% & 0.586 \\
Logic-Q w/o sketch                           & 4.1\% & 49.4\%& 21.1\% & -37.1\% & 0.366 & 5.9\% & 55.8\% & 32.5\% & -61.2\%& 0.416 \\
\rowcolor{mygray}
\textbf{Logic-Q (Ours)}        & \textcolor{red}{\textbf{11.8\%}}& \textcolor{red}{\textbf{109.3\%}} & 21.3\% & \textcolor{red}{\textbf{-31.9\%}}& \textcolor{red}{\textbf{0.624}} & \textcolor{red}{\textbf{24.1\%}} & \textcolor{red}{\textbf{411.7\%}} & 29.7\% & \textcolor{red}{\textbf{-30.9\%}} & \textcolor{red}{\textbf{1.135}}\\
\hline
{Improvement over SOTA }                        & {\textcolor{marine}{\textbf{122.6\%}}}                    & {\textcolor{marine}{\textbf{119.0\%}}}                &{\textbf{-}}             &  {\textcolor{marine}{\textbf{14.0\%}}}      & {\textcolor{marine}{\textbf{31.9\%}}}               & {\textcolor{marine}{\textbf{25.5\%}}}                 & {\textcolor{marine}{\textbf{75.5\%}}}                                       & {\textbf{-}}                & {\textcolor{marine}{\textbf{35.1\%}}}      & {\textcolor{marine}{\textbf{22.6\%}}}                                                  \\
\hline
\multicolumn{11}{c}{Cryptocurrency market}                                                                                                                                                                                                                                                                                                                                                                                                                       \\ 
\hline
\multirow{2}{*}{Methods}                 & \multicolumn{5}{c|}{Cash Trading}                                                                                                                                                           & \multicolumn{5}{c}{Margin Trading}                                                                                                                                                                                \\ 
\cmidrule{2-11}
                                         & AR                       & CR                   & AV                   & MD                    & SR                              & AR                       & CR                    & AV                   & MD                                          & SR                             \\ 
\midrule
BAH                                    & 28.7\%                              & 228.4\%                               & 65.7\%                             & -80.9\%                             & 0.718                              & 28.7\%                              & 228.4\%                                & 65.7\%                             & -80.9\%                                                   & 0.718                              \\
DDPG                                     & 28.5\%                        & 136.1\%                        & 68.7\%                     & -77.9\%                     & 0.709                        & 0\%                       & 0\%                       & 267.7\%                    & -100\%                                            & 0                      \\
A2C                                      & 25.3\%                       & 117.0\%         & 56.7\%                                   & -71.5\%                      & 0.682                       & 39.7\%                        & 212.3\%                        & 103.1\%                       & -87.9\%                                          & 0.844                      \\
PPO                                      & 46.7\%                       & 268.9\%                       & 64.8\%                    & -76.3\%                     & 0.915                       & 63.5\%                       & 433.8\%                       & 76.5\%                        & -88.3\%                                           & 1.020                        \\
AlphaMix                                      & 54.4\%                       & 339.1\%                       & 47.4\%                     & -50.3\%                    & 1.162                     & 87.4\%                        & 750.9\%                      & 71.5\%                        & -63.2\%                                           & 1.242                       \\

Sharpe-Ens                                 & 20.9\%                       & 91.1\%                           & 60.2\%                    & -69.6\%                       & 0.619                       & 44.1\%                     & 278.7.5\%                       & \textcolor{red}{\textbf{68.6\%}}                     & -79.5\%                                            & 0.842                       \\
AlphaMix (Aug)                                      & 56.4\%                        & 376.9\%                      & 49.8\%                     & -47.9\%                     & 1.157                      & 86.8\%                        & 736.4\%                      & 70.9\%                       & -65.1\%                                           & 1.230 \\
Sharpe-Ens (Aug)                                 & 22.7\%                        & 100.6\%                           & 60.6\%                     & -71.5\%                       & 0.643                       & 45.7\%                       & 301.5\%                       &69.1\%                      & -80.3\%                                            & 0.879                       \\
Logic-Q w/o sketch                           & 42.4\%                        & 264.7\%                      & 66.4\%                        & -71.3\%                     & 0.824                       & 75.9\%                       & 583.8\%                        & 75.4\%                    & -86.1\%                                          & 1.129                       \\
\rowcolor{mygray}
\textbf{Logic-Q (Ours)}        & \textcolor{red}{\textbf{63.6\%}}            & \textcolor{red}{\textbf{437.9\%}}                                              & \textcolor{red}{\textbf{42.7\%}}              & \textcolor{red}{\textbf{-46.1\%}}               & \textcolor{red}{\textbf{1.268}}                                         & \textcolor{red}{\textbf{92.9\%}}       & \textcolor{red}{\textbf{836.6\%}}   & 70.6\%    & \textcolor{red}{\textbf{-42.8\%}}                                                          & \textcolor{red}{\textbf{1.291}}                                                      \\
\hline
{Improvement over SOTA }                        & {\textcolor{marine}{\textbf{12.8\%}}}                    & {\textcolor{marine}{\textbf{16.2\%}}}                &{\textcolor{marine}{\textbf{10.0\%}}}            &  {\textcolor{marine}{\textbf{3.8\%}}}      & {\textcolor{marine}{\textbf{9.1\%}}}               & {\textcolor{marine}{\textbf{6.3\%}}}                 & {\textcolor{marine}{\textbf{13.6\%}}}                                       & {\textbf{-}}                   & {\textcolor{marine}{\textbf{56.4\%}}}      & {\textcolor{marine}{\textbf{4.0\%}}}                                                  \\
\bottomrule
\end{tabular}
}
\caption{The comparative results of different methods on the stock trading task.}
\label{tab:st}
\vspace{-1.5em}
\end{table*}
\subsubsection{Compared Methods}
\label{sec:st_baselines}
We compare our methods with the mainstream rule-based strategies, utilizing the Stock Market Index ($p_t$) as a widely accepted performance baseline. For the cryptocurrency sector, due to the lack of a universally accepted market index, we employ an equal-weight Buy and Hold (BAH) \cite{gort2022deep} strategy as a comparable benchmark. The DRL strategies we compare against include the state-of-the-art DRL strategies, namely DDPG~\cite{lillicrap2015continuous}, PPO~\cite{schulman2017proximal}, Sharpe-Ens~\cite{yang2020deep} and AlphaMix~\cite{sun2023mastering} (refer to the supplementary materials for the details of these methods).

\subsubsection{Evaluation Measurements} 
We evaluate the performance of the ST strategies with 5 different measurements: (1) Annualized return (AR): Annualized return is a measure of the average annual rate of return on an investment over a specified period, (2) Cumulative return (CR): Cumulative return is the total amount of return on an investment over a specific period, (3) Annualized volatility (AV): Annualized volatility is a measure of the degree of variation of an investment's returns over a specific period, expressed as an annualized percentage, (4) Maximum drawdown (MD): Maximum drawdown is a measure of the largest percentage decline in the value of an investment from its peak to its trough over a specific period, (5) Sharpe ratio (SR)~\cite{sharpe1994sharpe}: Sharpe ratio is a measure of risk-adjusted return that compares the excess return of an investment over the risk-free rate to its volatility. 

\subsubsection{Single-model RL Improvement (RQ1)}
We first evaluate whether Logic-Q can effectively improve the performance of a single-model RL policy on the order execution task. 
The results are shown in \Cref{tab:oe}. It can be observed that our method substantially increases total returns and reduces maximum drawdowns while maintaining volatility within an acceptable range. 
To demonstrate the imperativeness of the program sketch, we conduct an ablation study. Concretely, we remove the program sketch of Logic-Q, which degenerates the model into a simple policy with softmax temperature scaling using Bayesian optimization (denoted as Logic-Q w/o sketch). For a fair comparison, we use the same training setup for the ablation. \Cref{tab:oe} shows that compared with Logic-Q, the performance of Logic-Q w/o PS significantly drops by all metrics, which demonstrates that the program sketch is effective in embedding human expertise regarding market trend. Besides, to demonstrate the improvement is brought by the model design instead of merely the market information, we augment the state-of-the-art DRL baseline (OPD) by concatenating the market features with the input state (\ie~$[\mathbf{s}_t, \mathbf{T}^{mar}]$) and follow the same training setup, denoted as OPD (Aug) (as shown in \Cref{tab:oe}). The results validate that the market information alone is ineffective in boosting the model's performance.   
We further conduct an action analysis of different methods. \Cref{fig:oe_cs} shows the order execution (selling) details (\ie~to sell a specific number of shares) of different methods on 3 different assets of 3 different days. The colored lines represent the volume of shares traded (\ie~sold) by different strategies per minute, the grey line is the the market price $p_t$ of the target trading asset. We can observe that Logic-Q manages to capture better opportunities than all the compared baseline trading strategies. For example, in Case 1, Logic-Q can well identify an incoming market descend and sell the shares at the highest price of the day while the other strategies fail to do so and sell at much lower prices. 
\subsubsection{Ensemble RL Improvement (RQ2)}
We then study whether Logic-Q is effective in improving the performance of an ensemble RL policy. We conduct experiments on the stock trading task. The results on US and HK stock markets, as well as cryptocurrency market are shown in \Cref{tab:st}. We can see that compared with the state-of-the-art baselines (\ie~Sharpe-Ens and AlphaMix), Logic-Q manages to achieve a much higher return while decreasing the maximum drawdown, which results in its highest Sharpe ratio. Similar to the OE analysis, we conduct an ablation by removing the program sketch from Logic-Q (denoted as Logic-Q w/o PS), which degenerates the model into a simple bagging-based ensemble strategy using Bayesian Optimization for the ensemble weight tuning. 
The results in \Cref{tab:md_st} show that the performance of Logic-Q w/o PS drops significantly compared with Logic-Q, which demonstrates the imperativeness of the program sketch. We also evaluate the performance of the state-of-the-art baseline models with market information augmentation (\ie~AlphaMix (Aug) and Sharpe-Ens (Aug)). The results in \Cref{tab:md_st} show that augmenting the model input with market information is ineffective in improving the models' performance.
\Cref{tab:md_st} shows the maximum drawdown of $p_t$, Sharpe-Ens, AlphaMix, and Logic-Q during the two crashes. The results indicate that compared to the previous state-of-the-art methods, Logic-Q presents much lower drawdowns when encountering market crashes. 
\begin{figure}[!t] 
\centering 
\includegraphics[width=0.47\textwidth]{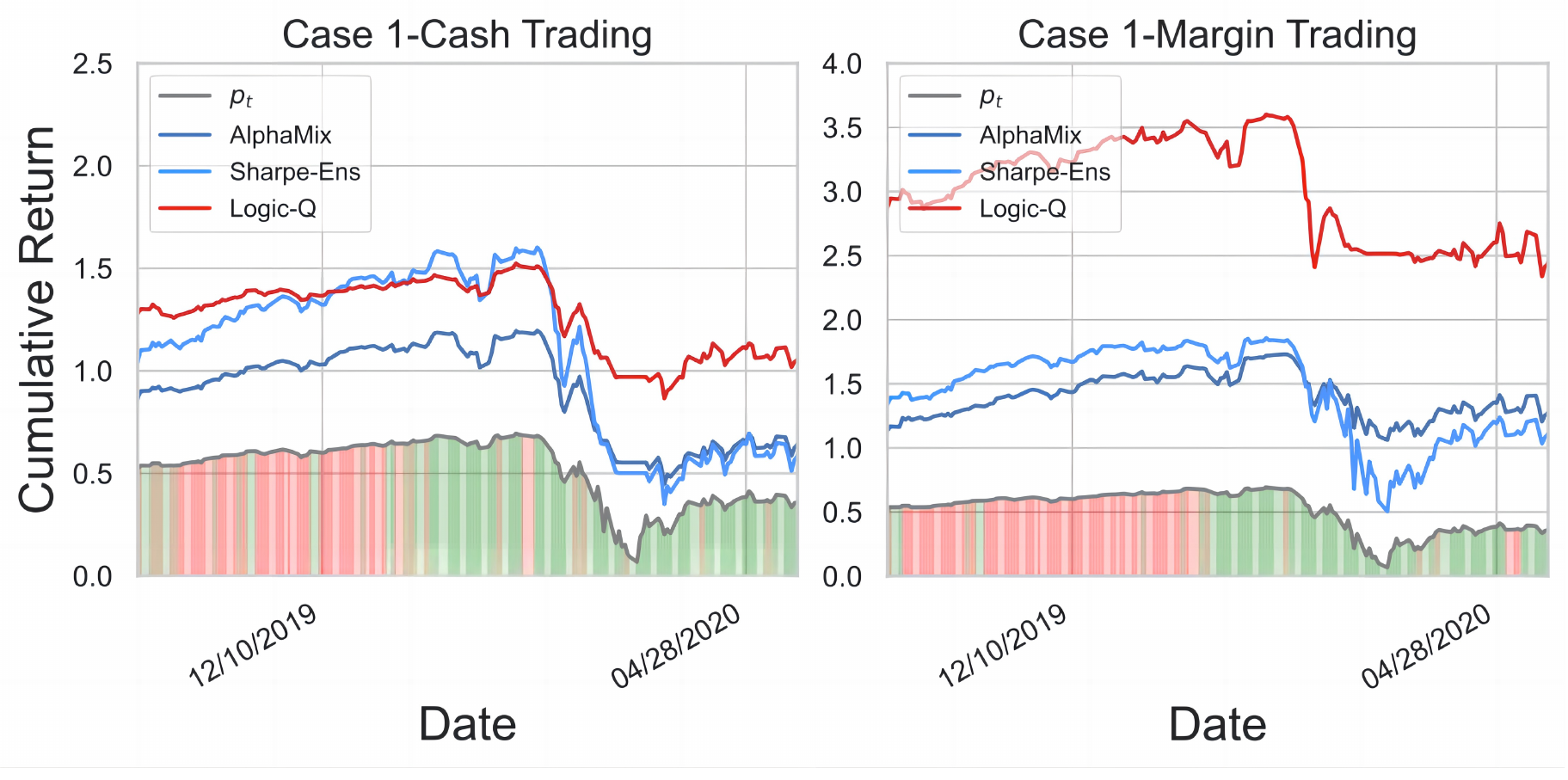} 
\caption{Cumulative return curve of different methods in the US market.} 
\vspace{-1em}
\label{fig:interpret} 
\end{figure}

\subsubsection{Interpretability of Program Sketch (RQ3)}
We further investigate how well the program sketch's market trend identification aligns with human expertise. We conduct the interpretability analysis based on significant market crash cases that happened from 03/09/2020 to 03/23/2020, due to the Fed's emergency rate cut that sparked a market-wide fire sale. The results are shown in \Cref{fig:interpret}.
Concretely, for each time step, if the optimized program sketch identifies the current market as an ascending market trend (\ie~rapid ascend or steady ascend), we denote it using a red vertical bar beneath
the market price curve $p_{t}$, and if the program sketch identifies the current market as a descending trend (\ie~rapid descend or steady descend), we denote it using a green bar. Otherwise, if the program sketch identifies the market as oscillation, we denote it using a white bar. We can observe that for the market crash under both the cash trading and margin trading settings, the area under the ascending period is reddish, and the optimized program sketch manages to timely foresee an incoming market fall (greenish area). The quantitative results are shown in \Cref{tab:md_st}, we can observe that Logic-Q manages to achieve much lower drawdown compared to the state-of-the-art AlphaMix and Sharpe-Ens baselines. In summary, the program sketch's market trend identification is interpretable by being strongly aligned with human judgment, which allows the Logic-Q model to be more risk-resilient by achieving much lower maximum drawdown under the market crash.

\section{Related Work}
\subsubsection{DRL for Finance}


The financial markets' dynamic nature and increasing volatility have exposed the limitations of traditional trading strategies, including autoregressive moving average (ARMA)~\cite{said1984testing}, pair trading~\cite{elliott2005pairs}, \etc. To tackle this challenge, the quantitative trading community is increasingly interested in utilizing deep learning and reinforcement learning techniques~\cite{sutton1998introduction,wang2019alphastock}, which have already achieved great performance in tackling complex stock trading~\cite{Wu2020AdaptiveST,lim2019enhancing}, order execution~\cite{yu2020reinforcement,breiman1996bagging}, and market making~\cite{zhao2021high,beysolow2019market}. 


\subsubsection{Program Synthesis by Sketching}
For many real-world problems, the program search space is intractable, which poses a great challenge for the synthesis model. \emph{Sketching}~\cite{solar2008program,singh2013automated,shah2020learning,medeiros2022can} is a novel program synthesis paradigm that proposes combining the human expert and the program synthesizer by embedding domain expert knowledge as general program sketches (\ie~a program with \emph{hole}). Then based on the program sketch, the synthesis is conducted to fill the \emph{hole}. In this way, the candidate program search space can be greatly reduced. Singh \etal~\cite{singh2013automated} propose a feedback generation system that automatically synthesizes program correction based on a general program sketch. 
\begin{table}[t]
\centering
\begin{tabular}{cc|c} 
\toprule
Method     & Case 1 CT  & Case 1 MT \\ 
\midrule
$p_t$       & -90.4\%  & -90.4\%   \\
AlphaMix  & -62.6\%  & -38.7\%   \\
Sharpe-Ens & -78.2 \%   & -72.8\%  \\
Logic-Q        & \textcolor{red}{\textbf{-43.3\%}} & \textcolor{red}{\textbf{-35.1\%}} \\

\bottomrule
\end{tabular}
\caption{Maximum drawdown of different methods during a market crash case. CT and MT denote cash trading and margin trading respectively.}
\label{tab:md_st}
\end{table}
\section{Conclusions}
In this paper, we propose a universal logic-guided DRL framework for quantitative trading, called program sketch-based tuning (Logic-Q), which embeds abstract human expert knowledge regarding market trends via the program synthesis by sketching paradigm. Logic-Q first introduces a general market trend-aware program sketch that describes the market in a logical manner while leaving numeric details that are hard to quantify as \emph{hole} to be optimized. Once parameterized, the program sketch is executed to determine the market trend of the current moment and the trained DRL strategy is then tuned accordingly without updating the neural network parameters. Finally, we introduce an effective optimization approach to optimize the symbolic program sketch. Experimental results validate that Logic-Q can significantly improve the performance of state-of-the-art DRL strategies while being extremely lightweight. 
\section{Acknowledgments}
This research / project is supported by the National Research Foundation, Singapore, and the Cyber Security Agency under its National Cybersecurity R\&D Programme (NCRP25-P04-TAICeN). It is also supported by DSO National Laboratories under the AI Singapore Programme (AISG Award No: AISG2-GC-2023-008), and the RIE2020 Industry Alignment Fund – Industry Collaboration Projects (IAF-ICP) Funding Initiative, as well as cash and in-kind contributions from the industry partner(s). Any opinions, findings and conclusions or recommendations expressed in this material are those of the author(s) and do not reflect the views of National Research Foundation, Singapore and Cyber Security Agency of Singapore. This work is funded by HKSAR RGC under Grant No. PolyU 15224823 and the Guangdong Basic and Applied Basic Research Foundation under Grant No. 2024A1515011524.

\bibliography{aaai25}

\begin{thebibliography}{42}
\providecommand{\natexlab}[1]{#1}

\bibitem[{Agarwala et~al.(2022)Agarwala, Schoenholz, Pennington, and Dauphin}]{agarwala2022temperature}
Agarwala, A.; Schoenholz, S.~S.; Pennington, J.; and Dauphin, Y. 2022.
\newblock Temperature check: theory and practice for training models with softmax-cross-entropy losses.
\newblock \emph{Transactions on Machine Learning Research}.

\bibitem[{Almgren and Chriss(2001)}]{almgren2001optimal}
Almgren, R.; and Chriss, N. 2001.
\newblock Optimal execution of portfolio transactions.
\newblock \emph{Journal of Risk}, 3: 5--40.

\bibitem[{Bertsimas and Lo(1998)}]{bertsimas1998optimal}
Bertsimas, D.; and Lo, A.~W. 1998.
\newblock Optimal control of execution costs.
\newblock \emph{Journal of financial markets}, 1(1): 1--50.

\bibitem[{Beysolow~II and Beysolow~II(2019)}]{beysolow2019market}
Beysolow~II, T.; and Beysolow~II, T. 2019.
\newblock Market making via reinforcement learning.
\newblock \emph{Applied Reinforcement Learning with Python: With OpenAI Gym, Tensorflow, and Keras}, 77--94.

\bibitem[{Breiman(1996)}]{breiman1996bagging}
Breiman, L. 1996.
\newblock Bagging predictors.
\newblock \emph{Machine learning}, 24: 123--140.

\bibitem[{Brock, Lakonishok, and LeBaron(1992)}]{brock1992simple}
Brock, W.; Lakonishok, J.; and LeBaron, B. 1992.
\newblock Simple technical trading rules and the stochastic properties of stock returns.
\newblock \emph{The Journal of finance}, 47(5): 1731--1764.

\bibitem[{Cao et~al.(2022)Cao, Li, Yang, Zhang, Zheng, Li, Hao, and Liu}]{cao2022galois}
Cao, Y.; Li, Z.; Yang, T.; Zhang, H.; Zheng, Y.; Li, Y.; Hao, J.; and Liu, Y. 2022.
\newblock GALOIS: boosting deep reinforcement learning via generalizable logic synthesis.
\newblock \emph{Advances in Neural Information Processing Systems}, 35: 19930--19943.

\bibitem[{Cartea, Jaimungal, and Penalva(2015)}]{cartea2015algorithmic}
Cartea, {\'A}.; Jaimungal, S.; and Penalva, J. 2015.
\newblock \emph{Algorithmic and high-frequency trading}.
\newblock Cambridge University Press.

\bibitem[{Cui et~al.(2023)Cui, Ding, Jin, and Zhang}]{cui2023portfolio}
Cui, T.; Ding, S.; Jin, H.; and Zhang, Y. 2023.
\newblock Portfolio constructions in cryptocurrency market: A CVaR-based deep reinforcement learning approach.
\newblock \emph{Economic Modelling}, 119: 106078.

\bibitem[{Ee et~al.(2020)Ee, Sharef, Yaakob, and Kasmiran}]{ee2020lstm}
Ee, Y.~K.; Sharef, N.~M.; Yaakob, R.; and Kasmiran, K.~A. 2020.
\newblock LSTM based recurrent enhancement of DQN for stock trading.
\newblock In \emph{2020 IEEE Conference on Big Data and Analytics (ICBDA)}, 38--44. IEEE.

\bibitem[{Elliott, Van Der~Hoek*, and Malcolm(2005)}]{elliott2005pairs}
Elliott, R.~J.; Van Der~Hoek*, J.; and Malcolm, W.~P. 2005.
\newblock Pairs trading.
\newblock \emph{Quantitative Finance}, 5(3): 271--276.

\bibitem[{Fang et~al.(2021)Fang, Ren, Liu, Zhou, Zhang, Bian, Yu, and Liu}]{fang2021universal}
Fang, Y.; Ren, K.; Liu, W.; Zhou, D.; Zhang, W.; Bian, J.; Yu, Y.; and Liu, T.-Y. 2021.
\newblock Universal trading for order execution with oracle policy distillation.
\newblock In \emph{Proceedings of the AAAI Conference on Artificial Intelligence}, volume~35, 107--115.

\bibitem[{Gort et~al.(2022)Gort, Liu, Sun, Gao, Chen, and Wang}]{gort2022deep}
Gort, B. J.~D.; Liu, X.-Y.; Sun, X.; Gao, J.; Chen, S.; and Wang, C.~D. 2022.
\newblock Deep reinforcement learning for cryptocurrency trading: Practical approach to address backtest overfitting.
\newblock \emph{arXiv preprint arXiv:2209.05559}.

\bibitem[{Guan and Liu(2021)}]{guan2021explainable}
Guan, M.; and Liu, X.-Y. 2021.
\newblock Explainable deep reinforcement learning for portfolio management: an empirical approach.
\newblock In \emph{Proceedings of the Second ACM International Conference on AI in Finance}, 1--9.

\bibitem[{Kakade et~al.(2004)Kakade, Kearns, Mansour, and Ortiz}]{kakade2004competitive}
Kakade, S.~M.; Kearns, M.; Mansour, Y.; and Ortiz, L.~E. 2004.
\newblock Competitive algorithms for VWAP and limit order trading.
\newblock In \emph{Proceedings of the 5th ACM conference on Electronic commerce}, 189--198.

\bibitem[{Lillicrap et~al.(2015)Lillicrap, Hunt, Pritzel, Heess, Erez, Tassa, Silver, and Wierstra}]{lillicrap2015continuous}
Lillicrap, T.~P.; Hunt, J.~J.; Pritzel, A.; Heess, N.; Erez, T.; Tassa, Y.; Silver, D.; and Wierstra, D. 2015.
\newblock Continuous control with deep reinforcement learning.
\newblock \emph{arXiv preprint arXiv:1509.02971}.

\bibitem[{Lim, Zohren, and Roberts(2019)}]{lim2019enhancing}
Lim, B.; Zohren, S.; and Roberts, S. 2019.
\newblock Enhancing time-series momentum strategies using deep neural networks.
\newblock \emph{The Journal of Financial Data Science}, 1(4): 19--38.

\bibitem[{Lin and Beling(2021)}]{lin2021end}
Lin, S.; and Beling, P.~A. 2021.
\newblock An end-to-end optimal trade execution framework based on proximal policy optimization.
\newblock In \emph{Proceedings of the Twenty-Ninth International Conference on International Joint Conferences on Artificial Intelligence}, 4548--4554.

\bibitem[{Liu et~al.(2022)Liu, Xia, Rui, Gao, Yang, Zhu, Wang, Wang, and Guo}]{liu2022finrl}
Liu, X.-Y.; Xia, Z.; Rui, J.; Gao, J.; Yang, H.; Zhu, M.; Wang, C.; Wang, Z.; and Guo, J. 2022.
\newblock FinRL-Meta: Market environments and benchmarks for data-driven financial reinforcement learning.
\newblock \emph{Advances in Neural Information Processing Systems}, 35: 1835--1849.

\bibitem[{Lo, Mamaysky, and Wang(2000)}]{lo2000foundations}
Lo, A.~W.; Mamaysky, H.; and Wang, J. 2000.
\newblock Foundations of technical analysis: Computational algorithms, statistical inference, and empirical implementation.
\newblock \emph{The journal of finance}, 55(4): 1705--1765.

\bibitem[{Markowitz(1952)}]{markowitz1952portfolio}
Markowitz, H. 1952.
\newblock Portfolio selection.
\newblock \emph{Journal of Finance}, 7(1): 77--91.

\bibitem[{Medeiros, Aleixo, and Lelis(2022)}]{medeiros2022can}
Medeiros, L.~C.; Aleixo, D.~S.; and Lelis, L.~H. 2022.
\newblock What can we learn even from the weakest? Learning sketches for programmatic strategies.
\newblock In \emph{Proceedings of the AAAI Conference on Artificial Intelligence}, volume~36, 7761--7769.

\bibitem[{Nan, Perumal, and Zaiane(2022)}]{nan2022sentiment}
Nan, A.; Perumal, A.; and Zaiane, O.~R. 2022.
\newblock Sentiment and knowledge based algorithmic trading with deep reinforcement learning.
\newblock In \emph{Database and Expert Systems Applications: 33rd International Conference, DEXA 2022, Vienna, Austria, August 22--24, 2022, Proceedings, Part I}, 167--180. Springer.

\bibitem[{Said and Dickey(1984)}]{said1984testing}
Said, S.~E.; and Dickey, D.~A. 1984.
\newblock Testing for unit roots in autoregressive-moving average models of unknown order.
\newblock \emph{Biometrika}, 71(3): 599--607.

\bibitem[{Schulman et~al.(2017)Schulman, Wolski, Dhariwal, Radford, and Klimov}]{schulman2017proximal}
Schulman, J.; Wolski, F.; Dhariwal, P.; Radford, A.; and Klimov, O. 2017.
\newblock Proximal policy optimization algorithms.
\newblock \emph{arXiv preprint arXiv:1707.06347}.

\bibitem[{Shah et~al.(2020)Shah, Zhan, Sun, Verma, Yue, and Chaudhuri}]{shah2020learning}
Shah, A.; Zhan, E.; Sun, J.; Verma, A.; Yue, Y.; and Chaudhuri, S. 2020.
\newblock Learning differentiable programs with admissible neural heuristics.
\newblock \emph{Advances in neural information processing systems}, 33: 4940--4952.

\bibitem[{Sharpe(1994)}]{sharpe1994sharpe}
Sharpe, W.~F. 1994.
\newblock The sharpe ratio, the journal of portfolio management.
\newblock \emph{Stanfold University, Fall}.

\bibitem[{Sharpe(1998)}]{sharpe1998sharpe}
Sharpe, W.~F. 1998.
\newblock The sharpe ratio.
\newblock \emph{Streetwise--the Best of the Journal of Portfolio Management}, 3: 169--185.

\bibitem[{Singh, Gulwani, and Solar-Lezama(2013)}]{singh2013automated}
Singh, R.; Gulwani, S.; and Solar-Lezama, A. 2013.
\newblock Automated feedback generation for introductory programming assignments.
\newblock In \emph{Proceedings of the 34th ACM SIGPLAN conference on Programming language design and implementation}, 15--26.

\bibitem[{Snoek, Larochelle, and Adams(2012)}]{snoek2012practical}
Snoek, J.; Larochelle, H.; and Adams, R.~P. 2012.
\newblock Practical bayesian optimization of machine learning algorithms.
\newblock \emph{Advances in neural information processing systems}, 25.

\bibitem[{Solar-Lezama(2008)}]{solar2008program}
Solar-Lezama, A. 2008.
\newblock \emph{Program synthesis by sketching}.
\newblock University of California, Berkeley.

\bibitem[{Stoll and Whaley(1990)}]{stoll1990dynamics}
Stoll, H.~R.; and Whaley, R.~E. 1990.
\newblock The dynamics of stock index and stock index futures returns.
\newblock \emph{Journal of Financial and Quantitative analysis}, 25(4): 441--468.

\bibitem[{Sun et~al.(2023)Sun, Wang, Xue, Lou, and An}]{sun2023mastering}
Sun, S.; Wang, X.; Xue, W.; Lou, X.; and An, B. 2023.
\newblock Mastering Stock Markets with Efficient Mixture of Diversified Trading Experts.

\bibitem[{Sutton, Barto et~al.(1998)}]{sutton1998introduction}
Sutton, R.~S.; Barto, A.~G.; et~al. 1998.
\newblock \emph{Introduction to reinforcement learning}, volume 135.
\newblock MIT press Cambridge.

\bibitem[{Verma et~al.(2018)Verma, Murali, Singh, Kohli, and Chaudhuri}]{verma2018programmatically}
Verma, A.; Murali, V.; Singh, R.; Kohli, P.; and Chaudhuri, S. 2018.
\newblock Programmatically interpretable reinforcement learning.
\newblock In \emph{International Conference on Machine Learning}, 5045--5054. PMLR.

\bibitem[{Wang et~al.(2019)Wang, Zhang, Tang, Wu, and Xiong}]{wang2019alphastock}
Wang, J.; Zhang, Y.; Tang, K.; Wu, J.; and Xiong, Z. 2019.
\newblock Alphastock: A buying-winners-and-selling-losers investment strategy using interpretable deep reinforcement attention networks.
\newblock In \emph{Proceedings of the 25th ACM SIGKDD international conference on knowledge discovery \& data mining}, 1900--1908.

\bibitem[{Wu et~al.(2020)Wu, Chen, Wang, Troiano, Loia, and Fujita}]{Wu2020AdaptiveST}
Wu, X.; Chen, H.; Wang, J.; Troiano, L.; Loia, V.; and Fujita, H. 2020.
\newblock Adaptive stock trading strategies with deep reinforcement learning methods.
\newblock \emph{Inf. Sci.}, 538: 142--158.

\bibitem[{Yang et~al.(2020)Yang, Liu, Zhong, and Walid}]{yang2020deep}
Yang, H.; Liu, X.-Y.; Zhong, S.; and Walid, A. 2020.
\newblock Deep reinforcement learning for automated stock trading: An ensemble strategy.
\newblock In \emph{Proceedings of the first ACM international conference on AI in finance}, 1--8.

\bibitem[{Yu et~al.(2020)Yu, Li, Chai, and Tang}]{yu2020reinforcement}
Yu, X.; Li, G.; Chai, C.; and Tang, N. 2020.
\newblock Reinforcement learning with tree-lstm for join order selection.
\newblock In \emph{2020 IEEE 36th International Conference on Data Engineering (ICDE)}, 1297--1308. IEEE.

\bibitem[{Zhang et~al.(2023)Zhang, Duan, Chen, Chen, Li, and Zhao}]{zhang2023towards}
Zhang, C.; Duan, Y.; Chen, X.; Chen, J.; Li, J.; and Zhao, L. 2023.
\newblock Towards generalizable reinforcement learning for trade execution.
\newblock In \emph{Proceedings of the Thirty-Second International Joint Conference on Artificial Intelligence}, 4975--4983.

\bibitem[{Zhao and Linetsky(2021)}]{zhao2021high}
Zhao, M.; and Linetsky, V. 2021.
\newblock High frequency automated market making algorithms with adverse selection risk control via reinforcement learning.
\newblock In \emph{Proceedings of the Second ACM International Conference on AI in Finance}, 1--9.

\bibitem[{Zhou et~al.(2019)Zhou, Liu, Siow, Du, and Liu}]{zhou2019devign}
Zhou, Y.; Liu, S.; Siow, J.; Du, X.; and Liu, Y. 2019.
\newblock Devign: Effective vulnerability identification by learning comprehensive program semantics via graph neural networks.
\newblock \emph{Advances in neural information processing systems}, 32.

\end{thebibliography}

\end{document}